\documentclass{elsartL}
\usepackage{graphicx}
\usepackage{amssymb}
\usepackage{amsmath}
\usepackage{appendix}
\begin{document}

\begin{frontmatter}
\title{The Kalman Filter: a didactical overview}
\author[IMS]{E.~Matsinos{$^*$}}
\address[IMS]{Institute of Mechatronic Systems, School of Engineering, Zurich University of Applied Sciences (ZHAW), Technikumstrasse 5, CH-8401 Winterthur, Switzerland}

\begin{abstract}
The present document aims at providing a short, didactical introduction to three standard versions of the Kalman filter, namely its variants identified as Basic, Extended, and Unscented. The application of 
these algorithms in three representative problems is discussed.
\end{abstract}
{$^*$}{E-mail: evangelos[DOT]matsinos[AT]sunrise[DOT]ch}
\end{frontmatter}

\section{\label{sec:Introduction}Introduction}

The seminal papers by Swerling \cite{swerl}, K\'alm\'an \cite{kal}, and K\'alm\'an and Bucy \cite{kalbuc} instigated the development of data-fusion algorithms for systems of known dynamical behaviour, the 
family of the Kalman Filters (KFs). The KFs are simple methods providing a trade-off between the expectations of the actual (also called `true') state of a system (obtained on the basis of a physical 
or mathematical model, to be called `dynamical' henceforth) and measurements yielding information on that state. It is assumed that the process of extracting information from the system does not disturb 
its temporal evolution. KFs have been applied in numerous research and development domains. Standard applications include the navigation and control of vehicles. Notable examples are the NASA Space Shuttle 
and the International Space Station. Other standard applications range from robotic motion and trajectory optimisation to machine learning. From the practical point of view, an advantage of the KFs is that 
their implementation in real-time applications does not require the storage of the history of the states of the system and of the measurements.

The application of KFs is usually thought of as comprising two phases, a categorisation which however rests upon a logistical basis: prediction and correction~\footnote{There is no consensus regarding the 
naming of these two phases.}. In the prediction phase, an estimate of the state of the system is obtained from the estimate of the system's last state. This prediction also takes account of the effects of 
the so-called control inputs, i.e., of the means to enforce changes onto the system via deliberate actions, e.g., as the case is when the brakes are applied in order to bring a moving vehicle to a halt at 
a road intersection. This prediction is usually referred to as the \emph{a priori} state estimate. In the correction phase, the \emph{a priori} state estimate is blended with the current-time measurement, 
to refine the system's state estimate. This improved estimate is called \emph{a posteriori} state estimate and represents a weighted average of the \emph{a priori} state estimate and of the measurement, a 
result which is more representative of the actual state of the system than either the \emph{a priori} state estimate or the measurement. Typically (but not necessarily), the prediction and correction 
phases alternate.

As aforementioned, the KFs make use of the state-transition model, the known control inputs, as well as the acquired measurements, to provide reliable estimates of the actual state of the system. One 
additional operator relates the actual state of the system and the result of the observation. The effects of the noise, in both phases, are included via appropriate covariance matrices. The Basic KF (BKF) 
is applicable in case of linear systems. Two other KFs have been developed in order to cover non-linear systems: the Extended KF (EKF) and the Unscented KF (UKF). The KFs do not rely on assumptions about 
the uncertainties involved in each particular problem. If the predictions of the state of a linear system and the measurements follow Gaussian distributions, the BKF is the optimal method for extracting 
the important information from the observations.

Aiming at providing simplified explanations of the essentials of the KFs, numerous papers have appeared in scientific and in popular journals. In my opinion, many of these papers do not serve their cause, 
due to the lack of clarity and/or of rigorousness, omission of examples, and, sadly, the mistakes they contain. Some of the available works give the impression that their authors' ambition was to provide 
recipes for fast implementation, rather than insight into the basic concepts underlying these algorithms. Rigorous and detailed overviews on the subject may be obtained from Refs.~\cite{wb,tbf}. The present 
document aims at providing a short, didactical introduction to three standard versions of the KF (i.e., the BKF, the EKF, and the UKF). It is addressed to the non-expert (I am no expert in Kalman filtering) 
who wishes to understand the basics of an algorithm prior to applying it in order to obtain solutions.

\section{\label{sec:Definitions}Definitions}

It is assumed that the temporal evolution of a deterministic system is under study and that measurements are performed on the system at specific time instants $t_0$, $t_1$, \dots, thus yielding $K$ time 
series, where $K$ stands for the dimension of the measurement vector $z$, i.e., for the number of independent pieces of information obtained from the system at one time instant. It is also assumed that the 
system is fully described (at all times) by a real $N$-dimensional (ket) state vector $x$ (e.g., encompassing position vectors of the system's components, velocities, etc.). Without loss of generality, 
one may choose $N$ to be the minimal number of quantities necessary for the full description of the system. Independent measurements imply that $K \leq N$. In the following, the state and measurement 
vectors at time instant $t_k$ will be denoted as $x_k$ and $z_k$, respectively. From now on, reference to the time instants $t_k$ will be made by simply quoting the index $k$.

The temporal evolution of a deterministic physical system is known, if the state of the system is \emph{exactly} known at one time instant. Of course, given that all measurements are subject to finite 
(non-zero) uncertainties, `exactitude' in experimental results is of no practical relevance. For systems which are observed only once and let evolve, the difference between the predicted and actual states 
is expected to increase with time; this is simply the consequence of error propagation in predictions. To obtain as reliable predictions as possible, it is required that the system be regularly monitored 
and its updated state be used in the derivation of new predictions. Of course, the frequency at which the state vector is updated is linked to the lapse of time within which the predictions may be considered 
as reliable.

\section{\label{sec:BKF}The Basic Kalman Filter}

The following matrices must be specified:
\begin{itemize}
\item $F_k$: the $N \times N$ state-transition matrix,
\item $H_k$: the $K \times N$ state-to-measurement transformation matrix,
\item $Q_k$: the $N \times N$ covariance matrix of the process noise, and
\item $R_k$: the $K \times K$ covariance matrix of the measurement noise.
\end{itemize}
One additional matrix ($B_k$, known as the control-input matrix) maps the control inputs onto the state of the system. This is an $N \times L$ matrix, where $L$ is the dimension of the (ket) control-input 
vector $u_k$.

Let us consider the problem of determining the position of a vehicle given the underlying deterministic dynamics, the effects of the application of the control inputs (accelerator, brakes, steering direction, 
etc.), and a time series of (noisy) measurements (of the position of the vehicle). Most vehicles are equipped with a GPS unit nowadays, providing an estimate of the vehicle's actual position within a few 
metres.

\begin{figure}
\begin{center}
\includegraphics[width=15.5cm]{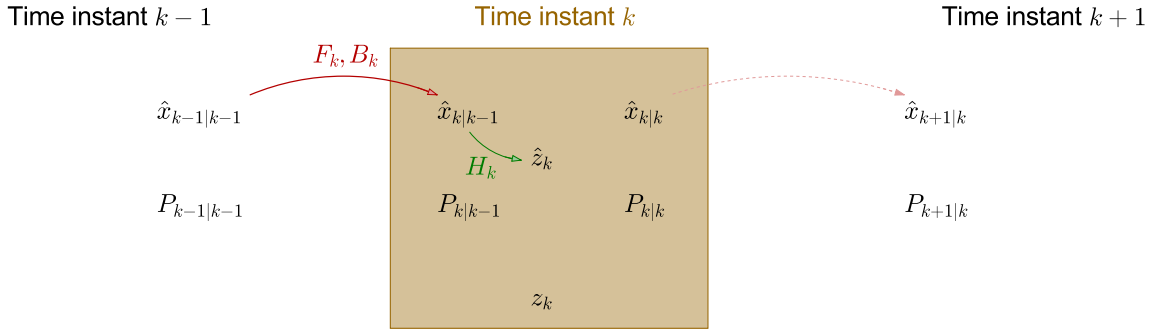}
\caption{\label{fig:BKFScheme}The scheme underlying the application of the Basic Kalman Filter. The output of the algorithm at time instant $k-1$ comprises estimates of the state vector $\hat{x}_{k-1 \vert k-1}$ 
and of the covariance matrix $P_{k-1 \vert k-1}$. Using the laws of motion ($F_k$) and the effects of the control inputs ($B_k$), one obtains predictions at time instant $k$, i.e., of the state vector 
$\hat{x}_{k \vert k-1}$ and of the covariance matrix $P_{k \vert k-1}$; these predictions comprise the input at that time instant. A prediction of the measurement at time instant $k$ is obtained from 
$\hat{x}_{k \vert k-1}$ on the basis of the state-to-measurement transformation matrix ($H_k$); this prediction is blended with the measurement $z_k$ at time instant $k$, resulting in updated predictions 
of the state vector $\hat{x}_{k \vert k}$ and of the covariance matrix $P_{k \vert k}$ at time instant $k$; these predictions comprise the output of the algorithm at time instant $k$. The next iteration 
involves quantities pertaining to time instants $k$ and $k+1$.}
\vspace{0.35cm}
\end{center}
\end{figure}

In the prediction phase, the laws of motion ($F_k$), using as input the vehicle's old position, and the changes induced by the control inputs ($u_k$) yield an \emph{a priori} state estimate of a new position 
at the time instant corresponding to the subsequent measurement. In the correction phase, a measurement of the vehicle's position $z_k$ is obtained from the GPS unit. The \emph{a priori} state estimate and 
the measurement are blended together, aiming at the minimisation of the difference between the \emph{a posteriori} state estimate and the actual position of the vehicle. The scheme is shown in 
Fig.~\ref{fig:BKFScheme}. The uncertainties, associated with the predictions and with the measurements, are taken into account via the covariance matrices $Q_k$ and $R_k$.

\subsection{\label{sec:BKFDetails}The details of the Basic Kalman Filter}

All KFs assume that the actual state of the system at time instant $k$ evolves from that at time instant $k-1$. In the BKF, the two actual states are related via the expression:
\begin{equation} \label{eq:EQ001}
x_k = F_k x_{k-1} + B_k u_k + q_k \, \, \, ,
\end{equation}
where $q_k$ is the process noise, drawn from a zero-mean multi-variate normal distribution with covariance matrix $Q_k$. At time instant $k$, a measurement $z_k$ is conducted, relating to the actual state 
of the system via the expression:
\begin{equation} \label{eq:EQ002}
z_k = H_k x_k + r_k \, \, \, ,
\end{equation}
where $r_k$ is the measurement noise, drawn from a zero-mean multi-variate normal distribution with covariance matrix $R_k$. At all time instants, the state estimates and the noise are not correlated. The 
matrices $F_k$ and $H_k$ are not dependent on the state vector of the system.

At time instant $k-1$, two quantities are obtained:
\begin{itemize}
\item $\hat{x}_{k \vert k-1}$: the \emph{a priori} state estimate and
\item $P_{k \vert k-1}$: the \emph{a priori} covariance matrix.
\end{itemize}
The subscript $n \vert m$ indicates estimates corresponding to time instant $n$, given information available at time instant $m$. At time instant $k$, two quantities are obtained:
\begin{itemize}
\item $\hat{x}_{k \vert k}$: the \emph{a posteriori} state estimate and
\item $P_{k \vert k}$: the \emph{a posteriori} covariance matrix.
\end{itemize}

{\bf Prediction equations}

The two prediction equations are:
\begin{itemize}
\item \emph{A priori} state estimate ($N$-dimensional ket)
\begin{equation} \label{eq:EQ003}
\hat{x}_{k \vert k-1} = F_k \hat{x}_{k-1 \vert k-1} + B_k u_k
\end{equation}
\item \emph{A priori} covariance matrix ($N \times N$ matrix)
\begin{equation} \label{eq:EQ004}
P_{k \vert k-1} = F_k P_{k-1 \vert k-1} F_k^T + Q_k
\end{equation}
\end{itemize}

{\bf Correction equations}

The correction scheme is based on the following relations:
\begin{itemize}
\item Predicted measurements ($K$-dimensional ket)
\begin{equation} \label{eq:EQ005}
\hat{z}_k = H_k \hat{x}_{k \vert k-1}
\end{equation}
\item Residuals ($K$-dimensional ket)
\begin{equation} \label{eq:EQ006}
\tilde{z}_k = z_k - \hat{z}_k
\end{equation}
\item Covariance matrix of the residuals ($K \times K$ matrix)
\begin{equation} \label{eq:EQ007}
S_k = H_k P_{k \vert k-1} H_k^T + R_k
\end{equation}
\item Optimal gain ($N \times K$ matrix)
\begin{equation} \label{eq:EQ008}
K_k = P_{k \vert k-1} H_k^T S_k^{-1}
\end{equation}
\item \emph{A posteriori} state estimate ($N$-dimensional ket)
\begin{equation} \label{eq:EQ009}
\hat{x}_{k \vert k} = \hat{x}_{k \vert k-1} + K_k \tilde{z}_k
\end{equation}
\item \emph{A posteriori} covariance matrix ($N \times N$ matrix) corresponding to the optimal gain
\begin{equation} \label{eq:EQ010}
P_{k \vert k} = (I_N - K_k H_k) P_{k \vert k-1}
\end{equation}
\end{itemize}
In the last expression, $I_N$ denotes the $N \times N$ identity matrix.

{\bf Some matrix algebra}

At first glance, relations (\ref{eq:EQ004},\ref{eq:EQ007},\ref{eq:EQ008},\ref{eq:EQ010}) may appear quasi-mystical, yet they are easily obtained on the basis of very simple mathematics. I start with the 
\emph{a priori} covariance matrix which is defined as the following expectation value:
\begin{equation} \label{eq:EQ011}
P_{k \vert k-1} = E \left[ (x_k - \hat{x}_{k \vert k-1}) (x_k - \hat{x}_{k \vert k-1})^T \right] \, \, \, .
\end{equation}
Invoking Eqs.~(\ref{eq:EQ001},\ref{eq:EQ003}), one obtains
\begin{align} \label{eq:EQ012}
P_{k \vert k-1} &= E \left[ \left( F_k (x_{k-1} - \hat{x}_{k-1 \vert k-1}) + q_k \right) \left( F_k (x_{k-1} - \hat{x}_{k-1 \vert k-1}) + q_k \right)^T \right]\nonumber\\
&= E \left[ \left( F_k (x_{k-1} - \hat{x}_{k-1 \vert k-1}) + q_k \right) \left( (x_{k-1} - \hat{x}_{k-1 \vert k-1})^T F_k^T + q_k^T \right) \right]\nonumber\\
&= E \left[ F_k (x_{k-1} - \hat{x}_{k-1 \vert k-1}) (x_{k-1} - \hat{x}_{k-1 \vert k-1})^T F_k^T + q_k q_k^T \right]\nonumber\\
&+ E \left[ F_k (x_{k-1} - \hat{x}_{k-1 \vert k-1}) q_k^T + q_k (x_{k-1} - \hat{x}_{k-1 \vert k-1})^T F_k^T \right]\nonumber\\
&= E \left[ F_k (x_{k-1} - \hat{x}_{k-1 \vert k-1}) (x_{k-1} - \hat{x}_{k-1 \vert k-1})^T F_k^T \right] + E \left[ q_k q_k^T \right]\nonumber\\
&= F_k P_{k-1 \vert k-1} F_k^T + Q_k \, \, \, ;
\end{align}
Equation (\ref{eq:EQ004}) is thus obtained. To derive this expression, use has been made of the assumption that the process noise is not correlated with the dynamics of the system.

I will next derive the expression for the covariance matrix of the residuals $S_k$. Obviously,
\begin{equation} \label{eq:EQ013}
S_k = E \left[ \tilde{z}_k \tilde{z}_k^T \right] \, \, \, ,
\end{equation}
which, after invoking Eqs.~(\ref{eq:EQ002},\ref{eq:EQ005},\ref{eq:EQ006}), attains the form:
\begin{align} \label{eq:EQ014}
S_k &= E \left[ \left( H_k (x_k - \hat{x}_{k \vert k-1}) + r_k \right) \left( H_k (x_k - \hat{x}_{k \vert k-1}) + r_k \right)^T \right]\nonumber\\
&= E \left[ \left( H_k (x_k - \hat{x}_{k \vert k-1}) + r_k \right) \left( (x_k - \hat{x}_{k \vert k-1})^T H_k^T + r_k^T \right) \right]\nonumber\\
&= E \left[ H_k (x_k - \hat{x}_{k \vert k-1}) (x_k - \hat{x}_{k \vert k-1})^T H_k^T + r_k r_k^T \right]\nonumber\\
&+ E \left[ H_k (x_k - \hat{x}_{k \vert k-1}) r_k^T + r_k (x_k - \hat{x}_{k \vert k-1})^T H_k^T \right]\nonumber\\
&= E \left[ H_k (x_k - \hat{x}_{k \vert k-1}) (x_k - \hat{x}_{k \vert k-1})^T H_k^T \right] + E \left[ r_k r_k^T \right]\nonumber\\
&= H_k P_{k \vert k-1} H_k^T + R_k \, \, \, ,
\end{align}
which is Eq.~(\ref{eq:EQ007}). To derive this expression, use has been made of the assumption that the measurement noise is not correlated with the system under observation.

Let me finally come to the derivation of the expression for the \emph{a posteriori} covariance matrix $P_{k \vert k}$, defined by the expression:
\begin{equation} \label{eq:EQ015}
P_{k \vert k} = E \left[ (x_k - \hat{x}_{k \vert k}) (x_k - \hat{x}_{k \vert k})^T \right] \, \, \, ,
\end{equation}
where the \emph{a posteriori} state estimate $\hat{x}_{k \vert k}$ is taken from Eq.~(\ref{eq:EQ009}). Invoking Eqs.~(\ref{eq:EQ002},\ref{eq:EQ005},\ref{eq:EQ006},\ref{eq:EQ007}), one obtains
\begin{align} \label{eq:EQ016}
P_{k \vert k} &= E \left[ \left( (I_N - K_k H_k)(x_k - \hat{x}_{k \vert k-1}) - K_k r_k \right) \left( (I_N - K_k H_k)(x_k - \hat{x}_{k \vert k-1}) - K_k r_k \right)^T \right]\nonumber\\
&= E \left[ \left( (I_N - K_k H_k)(x_k - \hat{x}_{k \vert k-1}) - K_k r_k \right) \left( (x_k - \hat{x}_{k \vert k-1})^T (I_N - H_k^T K_k^T) - r_k^T K_k^T \right) \right]\nonumber\\
&= E \left[ (I_N - K_k H_k)(x_k - \hat{x}_{k \vert k-1}) (x_k - \hat{x}_{k \vert k-1})^T (I_N - H_k^T K_k^T) \right]\nonumber\\
&+ E \left[ K_k r_k r_k^T K_k^T \right]\nonumber\\
&- E \left[ (I_N - K_k H_k)(x_k - \hat{x}_{k \vert k-1}) r_k^T K_k^T \right]\nonumber\\
&- E \left[ K_k r_k (x_k - \hat{x}_{k \vert k-1})^T (I_N - H_k^T K_k^T) \right]\nonumber\\
&= (I_N - K_k H_k) P_{k \vert k-1} (I_N - H_k^T K_k^T) + K_k R_k K_k^T \nonumber\\
&= P_{k \vert k-1} - P_{k \vert k-1} H_k^T K_k^T - K_k H_k P_{k \vert k-1} + K_k H_k P_{k \vert k-1} H_k^T K_k^T + K_k R_k K_k^T \nonumber\\
&= P_{k \vert k-1} - P_{k \vert k-1} H_k^T K_k^T - K_k H_k P_{k \vert k-1} + K_k S_k K_k^T \, \, \, ;
\end{align}
this relation is known as the `Joseph form'. The matrix $K_k$ may be chosen such that $P_{k \vert k}$ be minimised, in which case it satisfies the relation:
\begin{equation} \label{eq:EQ017}
K_k S_k = P_{k \vert k-1} H_k^T
\end{equation}
or, equivalently,
\begin{equation} \label{eq:EQ018}
K_k = P_{k \vert k-1} H_k^T S_k^{-1} \, \, \, .
\end{equation}
The quantity $K_k$ of Eqs.~(\ref{eq:EQ017},\ref{eq:EQ018}) is the so-called optimal gain. Substituting the optimal gain into Eq.~(\ref{eq:EQ016}) and using the fact that the covariance matrix $P_{k \vert k-1}$ 
is symmetric (hence its transpose is equal to the matrix itself), one finally obtains
\begin{equation} \label{eq:EQ019}
P_{k \vert k} = (I_N - K_k H_k) P_{k \vert k-1} \, \, \, .
\end{equation}
It should be stressed that Eq.~(\ref{eq:EQ019}) is applicable for the optimal gain, whereas Eq.~(\ref{eq:EQ016}) is the expression for the \emph{a posteriori} covariance matrix for arbitrary gain. The 
importance of the model predictions, relative to the measurements, is regulated by the gain matrix. Large values of the gain-matrix elements place larger weight on the measurements; for small values, 
the filtered data follow the model predictions more closely.

\subsection{\label{sec:BKFPerformance}Performance of the Basic Kalman Filter}

The BKF yields the optimal solution when the following conditions are fulfilled:
\begin{itemize}
\item the dynamics of the system is known and linear,
\item the noise is white and Gaussian, and
\item the covariance matrices of the noise ($Q_k$ and $R_k$) are known.
\end{itemize}

Experience shows that, in case of incomplete modelling of the dynamical behaviour of a system, the performance of the filter deteriorates. There are also issues relating to its numerical stability: round-off 
uncertainties may lead to noise covariance matrices which are not positive-definite, as the case is when they involve small values or when the noise level is high.

\subsection{\label{sec:BKFExample}Example of an application of the Basic Kalman Filter}

One representative example of the straightforward application of the BKF deals with the motion of a massive object in the gravitational field of the Earth; treated in several textbooks is the case where the 
air resistance is neglected. (For the sake of completeness, the dynamical model when the resistance is proportional to the velocity of the object is detailed in Appendix \ref{App:AppA}.) The object is released 
at height $x=x_0$ ($x=0$ on the surface of the Earth, positive upwards or, better expressed, away from the centre of gravity) at time $t=t_0$ and the release velocity is equal to $v_0$, in the radial 
direction of the gravitational field. Without loss of generality, $t_0$ may be fixed to $0$ (i.e., the clock starts ticking at the moment the object is released). Elementary Physics dictates that the height 
$x(t)$ and the velocity $v(t)$ of the object are given by the expressions:
\begin{equation} \label{eq:EQ020}
x(t) = x_0 + v_0 t - \frac{1}{2} g t^2
\end{equation}
and
\begin{equation} \label{eq:EQ021}
v(t) = v_0 - g t \, \, \, ,
\end{equation}
where $g=9.80665$ m s$^{-2}$ \cite{pdg} is the standard gravitational acceleration. Of course, these two relations are valid so long as $g$ may be assumed to be constant. Discretising the motion in the 
time domain, one may put these two equations into the compact form:
\begin{equation} \label{eq:EQ022}
\left(
\begin{array}{c}
x_k\\
v_k
\end{array} \right)
= \left(
\begin{array}{cc}
1 & \Delta t_{k-1}\\
0 & 1
\end{array} \right)
\left(
\begin{array}{c}
x_{k-1}\\
v_{k-1}
\end{array} \right)
-g \left(
\begin{array}{c}
\Delta t_{k-1}^2/2\\
\Delta t_{k-1}
\end{array} \right)
\end{equation}
where $\Delta t_{k-1}=t_k-t_{k-1}$. This simple deterministic system is fully described if the position and the velocity of the object are known at (any) one time instant. A comparison of this expression 
with Eq.~(\ref{eq:EQ001}) in the absence of process noise ($q_k=0$) leads to the straightforward identification of the state-transition and control-input matrices, as well as of the control-input vector 
$u_k$ which reduces (in this problem) to a mere constant ($-g$). In the absence of a control input, no force is exerted on the object and Newton's first law of motion is recovered. The dimensions $N$ and 
$L$ are equal to $2$ and $1$, respectively.

\begin{figure}
\begin{center}
\includegraphics[width=15.5cm]{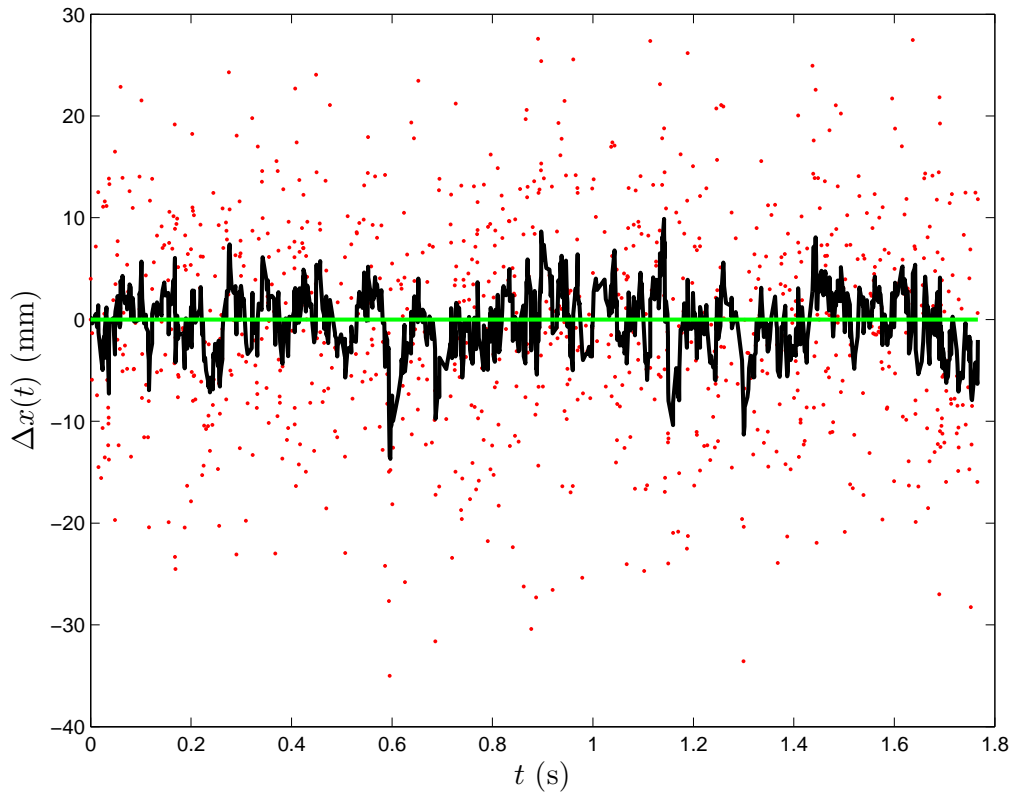}
\caption{\label{fig:Altitude}Difference of the position (height) to the exact solution corresponding to the free-fall example of Section \ref{sec:BKFExample}: the raw data are represented by the points, 
the filtered data by the black zigzag line. The green line marks the level of the exact noise-free solution.}
\vspace{0.35cm}
\end{center}
\end{figure}

\begin{figure}
\begin{center}
\includegraphics[width=15.5cm]{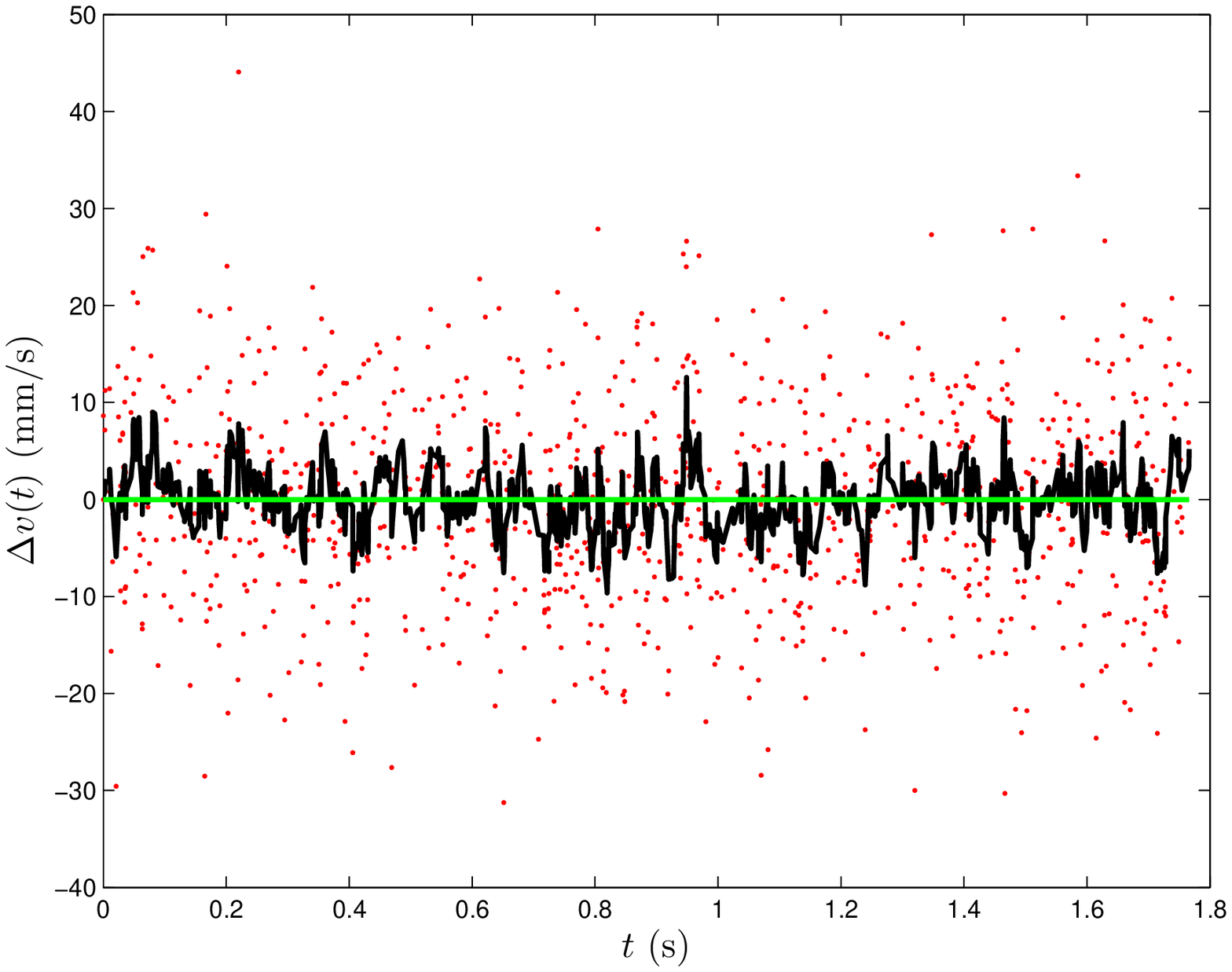}
\caption{\label{fig:Velocity}Same as Fig.~\ref{fig:Altitude} for the velocity.}
\vspace{0.35cm}
\end{center}
\end{figure}

Using the initial conditions of $x_0 = 10$ m and $v_0 = 3$ m/s, adding white Gaussian noise to the data (process: $2$ mm and $2$ mm/s; measurement: $10$ mm and $10$ mm/s), and applying the filter to $1\,000$ 
measurements of the height and velocity ($K=2$) for $H_k=I_2$, one obtains Figs.~\ref{fig:Altitude} and \ref{fig:Velocity}. Evidently, the filter is successful in reducing the noise present in the raw 
measurements.

\begin{figure}
\begin{center}
\includegraphics[width=15.5cm]{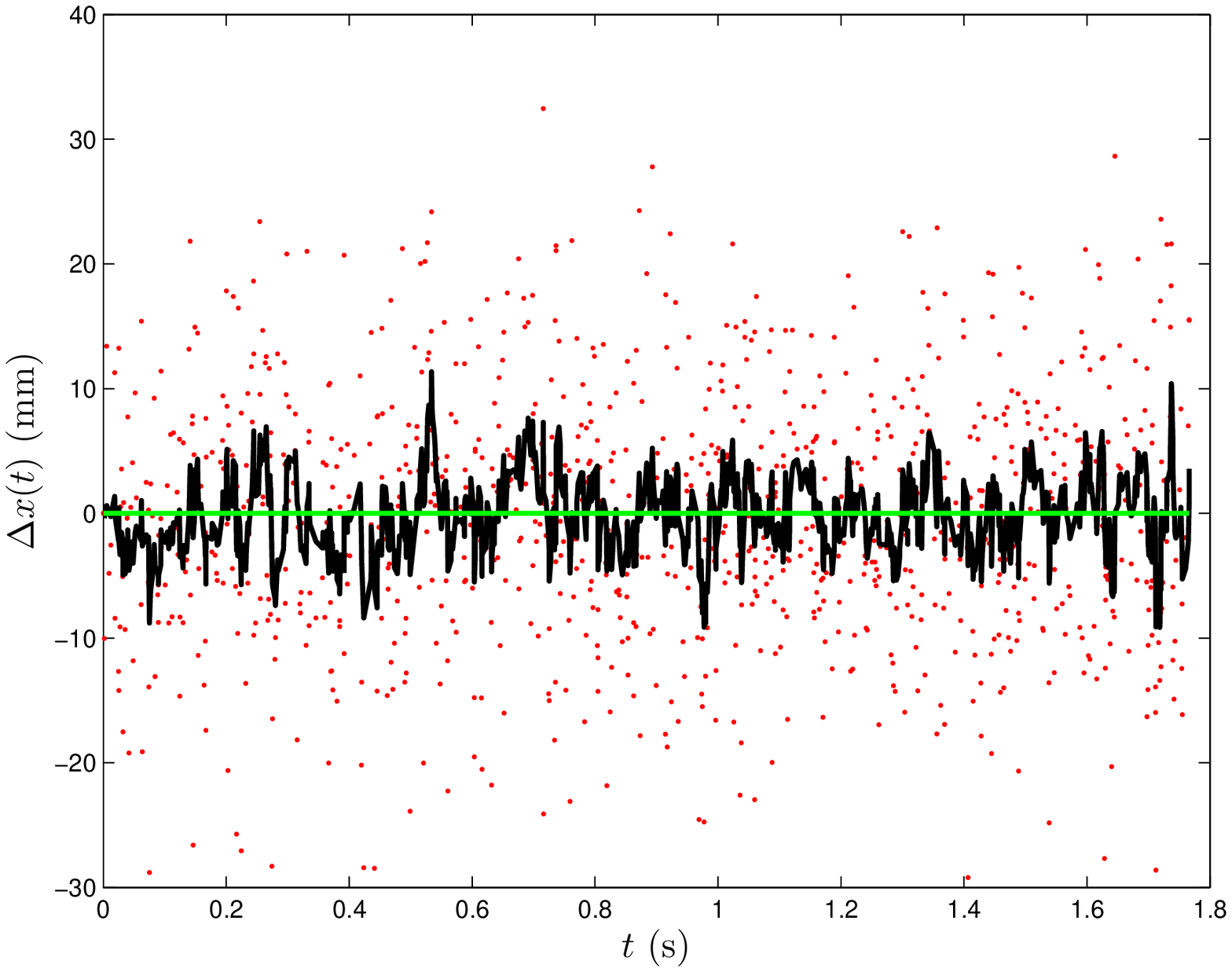}
\caption{\label{fig:AltitudeO1M}Same as Fig.~\ref{fig:Altitude} in the case that only the height of the object is measured ($K=1$).}
\vspace{0.35cm}
\end{center}
\end{figure}

\begin{figure}
\begin{center}
\includegraphics[width=15.5cm]{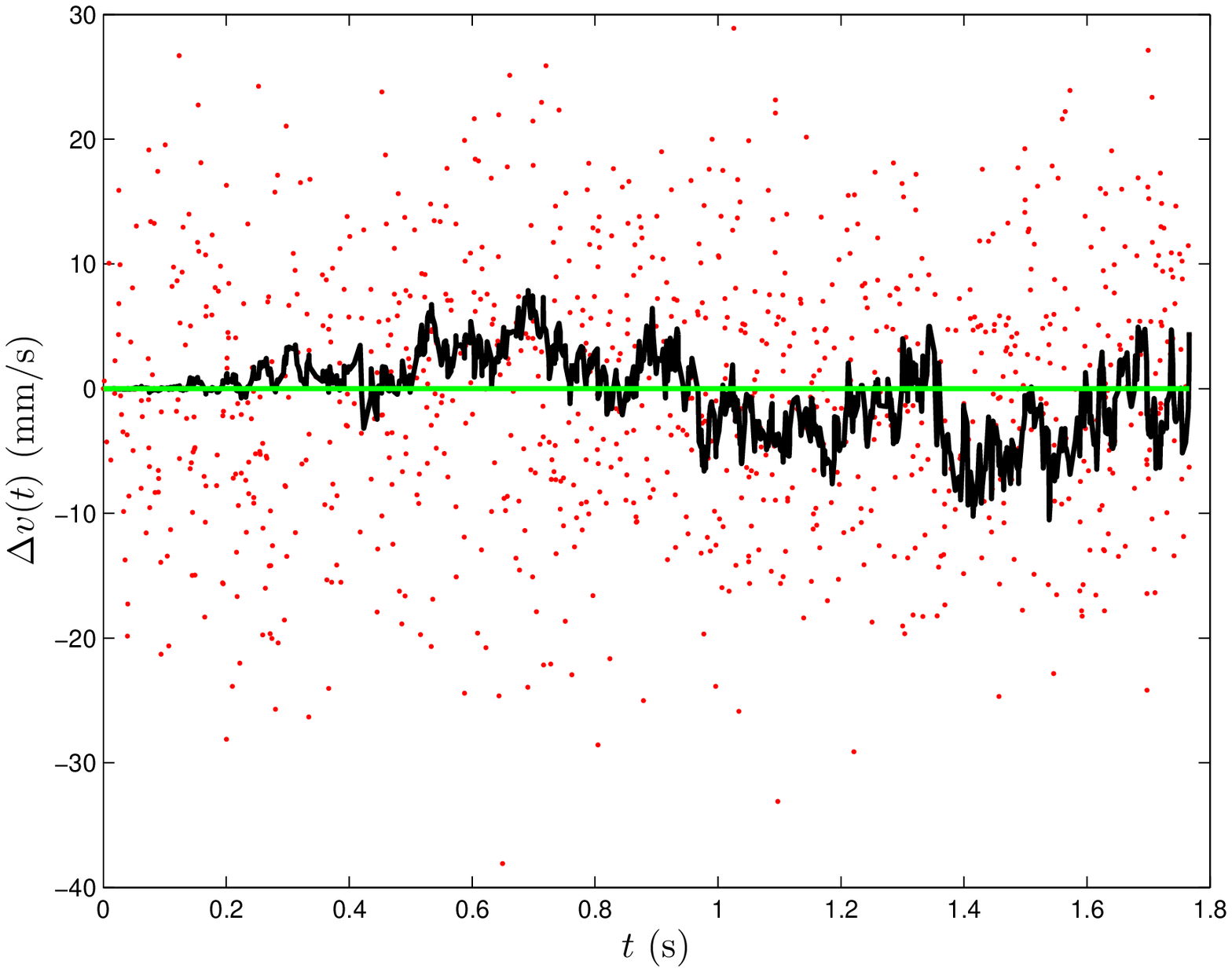}
\caption{\label{fig:VelocityO1M}Same as Fig.~\ref{fig:Velocity} in the case that only the height of the object is measured ($K=1$).}
\vspace{0.35cm}
\end{center}
\end{figure}

In comparison, the time dependence of the height and of the velocity is also shown in case that only the height is monitored ($K=1$), see Figs.~\ref{fig:AltitudeO1M} and \ref{fig:VelocityO1M} for the 
same initial conditions as before. Evidently, the filter is successful in reducing the noise present in the measurements of the height. On the other hand, some systematic effects are seen in 
Fig.~\ref{fig:VelocityO1M}.

At this point, one remark needs to be made. The BKF reduces the noise level; it does not eliminate it. This implies that, if a smooth solution is sought in an application, a smoothing procedure must be 
applied to the filtered data. In such a case, the process of applying the BKF to the raw measurements might have been an unnecessary step; an optimisation scheme (i.e., directly fitting to the measurements) 
might have been a more appropriate approach.

\section{\label{sec:EKF}The Extended Kalman Filter}

In the EKF, the state-transition matrix $F$ and the state-to-measurement transformation matrix $H$ are replaced by differentiable functions, to be denoted by $f$ and $h$, respectively. I will assume that 
the function $f$ also includes the effects of the control inputs. Equations (\ref{eq:EQ001},\ref{eq:EQ002}) take the form:
\begin{equation} \label{eq:EQ023}
x_k = f(x_{k-1},u_k) + q_k
\end{equation}
and
\begin{equation} \label{eq:EQ024}
z_k = h(x_k) + r_k \, \, \, .
\end{equation}

To make use of the prediction and correction expressions developed in the case of the BKF, i.e., of the scheme outlined by Eqs.~(\ref{eq:EQ003}-\ref{eq:EQ010}), the functions $f$ and $h$ are linearised 
around the state estimate, and the corresponding Jacobian matrices of the transformations $x_{k-1} \to x_k$ and $x_k \to z_k$ are analytically evaluated and supplied to the algorithm. Equations 
(\ref{eq:EQ003},\ref{eq:EQ005}) take the form:
\begin{equation} \label{eq:EQ025}
\hat{x}_{k \vert k-1} = f(\hat{x}_{k-1 \vert k-1},u_k)
\end{equation}
and
\begin{equation} \label{eq:EQ026}
\hat{z}_k = h(\hat{x}_{k \vert k-1}) \, \, \, .
\end{equation}
The matrices $F_k$ and $H_k$ in Eqs.~(\ref{eq:EQ004},\ref{eq:EQ007}-\ref{eq:EQ010}) must be replaced by the corresponding Jacobians.

\subsection{\label{sec:EKFExample}Example of an application of the Extended Kalman Filter}

The Lotka-Volterra equations~\footnote{The first version of these equations appeared in a paper by Lotka in the beginning of the $20^{\rm th}$ century \cite{lot}.} are non-linear, first-order, differential 
equations, which are frequently employed in the modelling of the dynamics of isolated ecosystems, consisting of a number of interacting (in terms of their dietary habits) biological species. The simplest 
of these ecosystems involves just two species: one serving as prey (herbivores), another as predators (preying on the herbivores). The temporal evolution of such a two-species ecosystem is deterministic 
and may be modelled according to the following set of equations.
\begin{align} \label{eq:EQ027}
\frac{dx}{dt} &= x (\alpha - \beta y)\nonumber\\
\frac{dy}{dt} &= y (-\gamma + \delta x) \, \, \, ,
\end{align}
where $x$ and $y$ denote the prey and predator populations, respectively; $\alpha$, $\beta$, $\gamma$, and $\delta$ are positive parameters. A numerical solution to this set of equations may be obtained with 
standard solvers, e.g., see Ref.~\cite{ptvf} (Chapter 17).

In the absence of predators ($y=0$), the first equation yields an exponentially growing prey population (which, of course, is unrealistic as vegetation will become scarce at some time, resulting in the 
decrease of the population). The decrease of the prey population due to predation is assumed to be proportional to the probability at which predators and prey occupy the same regions of spacetime; this 
probability involves the product $x y$. On the other hand, in the absence of prey ($x=0$), the predator population is bound to decrease exponentially with time; this is taken into account by the negative 
sign of the term $\gamma y$ in the second of Eqs.~(\ref{eq:EQ027}). Finally, the reproduction rate of the predators is related to the availability of food, i.e., to the probability (once again) of the 
encounters between predators and prey (term $\delta x y$).

\begin{figure}
\begin{center}
\includegraphics[width=15.5cm]{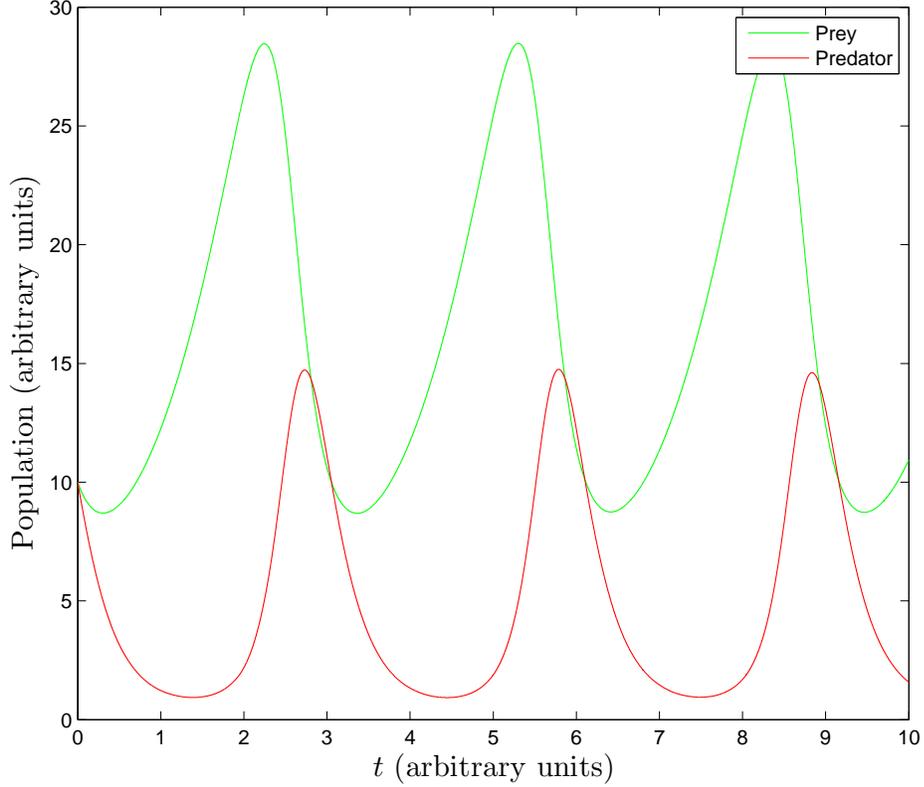}
\caption{\label{fig:PreyPredator}One example of the temporal evolution of a two-species ecosystem. The parameters $\alpha$, $\beta$, $\gamma$, and $\delta$ were fixed to $1.0$, $0.2$, $5.0$, and $0.3$, 
respectively. The initial conditions for the populations were: $x(0)=y(0)=10$.}
\vspace{0.35cm}
\end{center}
\end{figure}

One example of the temporal evolution of a two-species ecosystem is illustrated in Fig.~\ref{fig:PreyPredator}. As expected, the two populations are intricately interrelated. With abundant prey, the predator 
population rises; this leads to the decrease of the prey population; this leads the decrease of the predator population; this leads to the increase of the prey population, and so on. One observes that the 
two scatter plots $x(t)$ and $y(t)$ are somewhat shifted (in time) with respect to each other, the prey-population maximum preceding that of the predators.

From Eqs.~(\ref{eq:EQ027}), one obtains the approximate expressions:
\begin{align} \label{eq:EQ028}
x_k &= x_{k-1} + x_{k-1} (\alpha - \beta y_{k-1}) \Delta t_{k-1}\nonumber\\
y_k &= y_{k-1} + y_{k-1} (-\gamma + \delta x_{k-1}) \Delta t_{k-1} \, \, \, ,
\end{align}
yielding the Jacobian matrix:
\begin{equation} \label{eq:EQ029}
F_{x_k/x_{k-1}} =
\left(
\begin{array}{cc}
1 + \alpha \Delta t_{k-1} - \beta y_{k-1} \Delta t_{k-1} & - \beta x_{k-1} \Delta t_{k-1}\\
\delta y_{k-1} \Delta t_{k-1} & 1 - \gamma \Delta t_{k-1} + \delta x_{k-1} \Delta t_{k-1}
\end{array} \right)
\end{equation}

\begin{figure}
\begin{center}
\includegraphics[width=15.5cm]{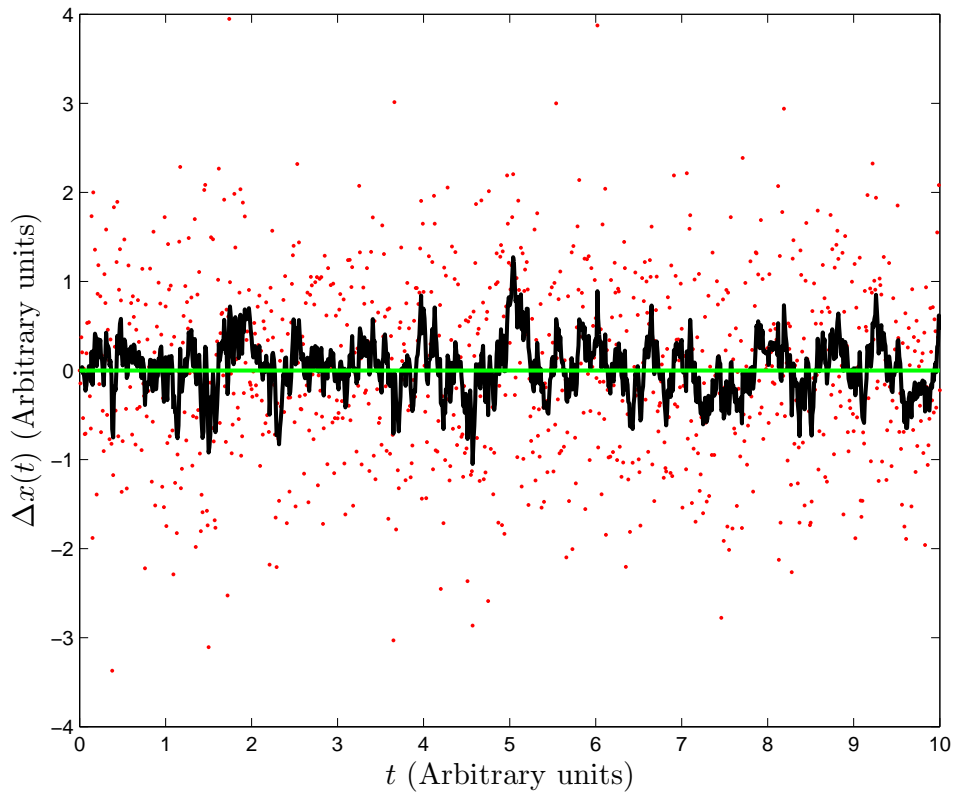}
\caption{\label{fig:Prey}Difference of the solution for the prey population to the numerical solution of the set of Eqs.~(\ref{eq:EQ027}), used in the modelling of the predator-prey problem of Section 
\ref{sec:EKFExample}: the raw data are represented by the points, the filtered data by the black zigzag line. The green line marks the level of the exact, noise-free solution.}
\vspace{0.35cm}
\end{center}
\end{figure}

\begin{figure}
\begin{center}
\includegraphics[width=15.5cm]{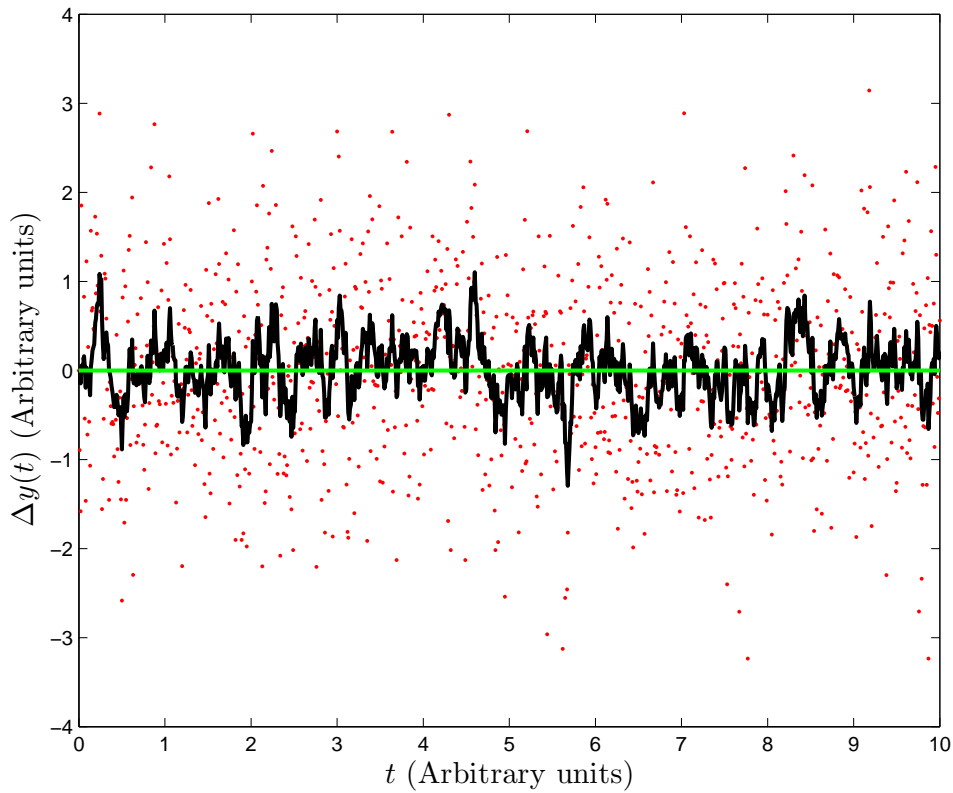}
\caption{\label{fig:Predator}Same as Fig.~\ref{fig:Prey} for the predator population.}
\vspace{0.35cm}
\end{center}
\end{figure}

Using the initial conditions of $x_0 = y_0 = 10$, adding white Gaussian noise to the data (process: $0.2$; measurement: $1.0$), and applying the filter to $1\,000$ measurements of the two populations 
($K=2$) for $H_k=I_2$, one obtains Figs.~\ref{fig:Prey} and \ref{fig:Predator}.

\section{\label{sec:UKF}The Unscented Kalman Filter}

It has been demonstrated that the EKF performs poorly when the functions $f$ and $h$ of Eqs.~(\ref{eq:EQ023},\ref{eq:EQ024}) are highly non-linear \cite{ju1,ju2,wm,kfi}. The selective (deterministic) 
sampling of these functions, rather than the use of single points in the derivation of predictions, has been put forth as the `treatment plan' for the linearisation problem: the UKF was thus established. 
One frequently advertised feature of this algorithm is that no Jacobian matrices need to be evaluated. In my opinion, the emphasis which many authors place on this `advantage' is misleading and 
counter-productive, as it diverts attention away from the main feature of the UKF, which is the inclusion of higher-order effects in the Taylor expansions of the non-linear functions $f$ and $h$.

\begin{figure}
\begin{center}
\includegraphics[width=15.5cm]{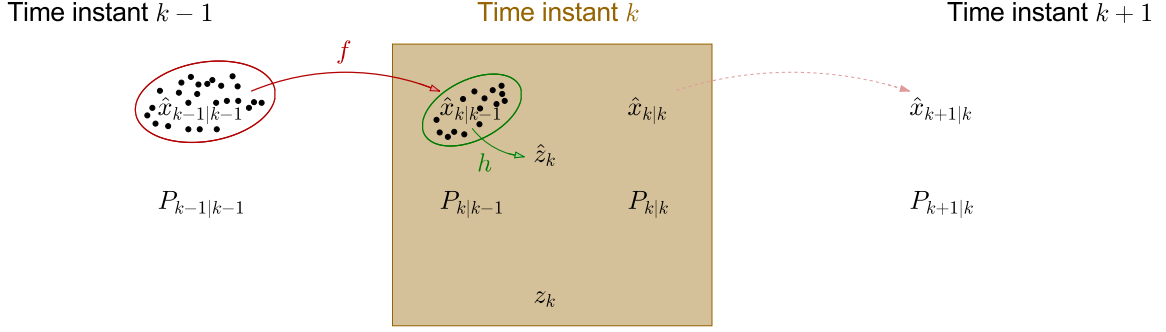}
\caption{\label{fig:UKFScheme}The adaptation of Fig.~\ref{fig:BKFScheme} for the Unscented Kalman Filter. The essential differences between the two figures pertain to the use of the functions $f$ and $h$ 
in the mappings $x_{k-1} \to x_k$ and $x_k \to z_k$, respectively, as well as to the introduction of $2 N + 1$ representative points (i.e., of the sigma points) in the derivation of the predictions.}
\vspace{0.35cm}
\end{center}
\end{figure}

I will next outline the procedure of applying the UKF, starting from the estimates $\hat{x}_{k-1 \vert k-1}$ and $P_{k-1 \vert k-1}$. Compact implementations of the UKF, inspired by Refs.~\cite{ju1,ju2,wm}, 
may be found in the literature, e.g., see Ref.~\cite{mw} (and several other later works); most of these implementations rest upon operations with enhanced state vector and covariance matrices, after the 
incorporation of the effects (i.e., of the expectation values and covariance matrices) of the process and/or of the measurement noise. However, I refrain from sacrificing straightforwardness for elegance, 
and describe herein only what is known as `standard' implementation of the UKF. A commendable effort towards explaining the application of the UKF to continuous-time filtering problems may be found in 
Ref.~\cite{sarkka}. A schematic overview of the UKF is given in Fig.~\ref{fig:UKFScheme}.

In the prediction phase, one first selects $2 N + 1$ representative points around $\hat{x}_{k-1 \vert k-1}$ (e.g., the so-called sigma points), propagates these points through the non-linear function $f$, 
and obtains estimates for $\hat{x}_{k \vert k-1}$ and $P_{k \vert k-1}$ as weighted averages over the transformed points. These estimates are expected to perform better than their EKF counterparts (which 
are solely based on one single point, namely $\hat{x}_{k-1 \vert k-1}$), the improvement originating in the inclusion of the higher-order effects in the mapping of the function $f$ (and, in the correction 
phase, of $h$). There is no unique prescription for drawing the sigma points. One of the possibilities, utilising two sigma points per component of the vector $\hat{x}_{k-1 \vert k-1}$ (as well as the 
central value $\hat{x}_{k-1 \vert k-1}$), is outlined as follows.
\begin{align} \label{eq:EQ030}
\bar{x}^0_{k-1 \vert k-1} &= \hat{x}_{k-1 \vert k-1}\nonumber\\
\bar{x}^i_{k-1 \vert k-1} &= \hat{x}_{k-1 \vert k-1} + \mathcal{X}_i \text{, for $1 \leq i \leq N$}\nonumber\\
\bar{x}^i_{k-1 \vert k-1} &= \hat{x}_{k-1 \vert k-1} - \mathcal{X}_{i-N} \text{, for $N + 1 \leq i \leq 2 N$}
\end{align}
The quantity $\mathcal{X}_j$ in Eqs.~(\ref{eq:EQ030}) represents the $j^{\rm th}$ column of the `square root' of the matrix
\begin{equation*}
(N + \lambda) P_{k-1 \vert k-1} \, \, \, ,
\end{equation*}
where $\lambda$ will be dealt with shortly. The Cholesky decomposition is a robust method for obtaining the `square root' matrix, e.g., see Ref.~\cite{ptvf} (Chapter 2.9). The sigma points come with two 
types of weights: one set ($w^i_x$) for the determination of state and measurement predictions, another ($w^i_p$) for the various covariance matrices relating to the method. These weights are defined as 
follows.
\begin{align} \label{eq:EQ031}
w^0_x &= \frac{\lambda}{N + \lambda}\nonumber\\
w^0_p &= w^0_x + 1 - \alpha^2 + \beta\nonumber\\
w^i_x &= w^i_p = \frac{1}{2(N + \lambda)}
\end{align}
The weights $w^i_x$ are normalised:
\begin{equation*}
\sum_{i=0}^{2 N} w^i_x = 1 \, \, \, .
\end{equation*}
The scaling parameter $\lambda$ is expressed in terms of two parameters, $\alpha \in (0,1]$ and $\kappa$, according to the formula:
\begin{equation} \label{eq:EQ032}
\lambda = \alpha^2 (N + \kappa) - N \, \, \, .
\end{equation}
Obviously, three parameters need adjustment before applying the UKF: $\alpha$, $\beta$, and $\kappa$. The parameters $\alpha$ and $\kappa$ regulate the range of values of the sigma points: the distance 
between corresponding sigma points widens with increasing values of these two parameters. A general method for obtaining the optimal values of $\alpha$, $\beta$, and $\kappa$ is still to be established; 
these parameters may be `fine-tuned', aiming at the optimisation of the results in each particular problem. In case of Gaussian distributions, the choice $\beta = 2$ is optimal \cite{wm}.

In the correction phase, one selects a new set of $2 N + 1$ representative points around $\hat{x}_{k \vert k-1}$ (taking the updated covariance matrix $P_{k \vert k-1}$ into account, in accordance with 
Eq.~(\ref{eq:EQ005})), propagates these points through the non-linear function $h$, and obtains estimates for $\hat{x}_{k \vert k}$ and $P_{k \vert k}$ as weighted averages over the transformed points. 
Two sigma points per component of the vector $\hat{x}_{k \vert k-1}$ (as well as the central value), are selected.
\begin{align} \label{eq:EQ033}
\bar{y}^0_{k \vert k-1} &= \hat{x}_{k \vert k-1}\nonumber\\
\bar{y}^i_{k \vert k-1} &= \hat{x}_{k \vert k-1} + \mathcal{Y}_i \text{, for $1 \leq i \leq N$}\nonumber\\
\bar{y}^i_{k \vert k-1} &= \hat{x}_{k \vert k-1} - \mathcal{Y}_{i-N} \text{, for $N + 1 \leq i \leq 2 N$}
\end{align}
The quantity $\mathcal{Y}_j$ represents the $j^{\rm th}$ column of the `square root' of the matrix
\begin{equation*}
(N + \lambda) P_{k \vert k-1} \, \, \, .
\end{equation*}
Apart from the use of a single point to utilising a set of appropriately selected points, one additional modification over the steps outlined by Eqs.~(\ref{eq:EQ005}-\ref{eq:EQ010}) is worth mentioning: 
in the UKF, the optimal gain matrix $K_k$ involves the so-called state-measurement cross-covariance matrix; the product $P_{k \vert k-1} H_k^T$ of Eq.~(\ref{eq:EQ008}) is replaced by this matrix.

{\bf Prediction equations}

The two prediction equations take the form:
\begin{itemize}
\item \emph{A priori} state estimate ($N$-dimensional ket)
\begin{equation} \label{eq:EQ034}
\hat{x}_{k \vert k-1} = \sum^{2 N}_{i=0} w^i_x f(\bar{x}^i_{k-1 \vert k-1})
\end{equation}
\item \emph{A priori} covariance matrix ($N \times N$ matrix)
\begin{equation} \label{eq:EQ035}
P_{k \vert k-1} = \sum^{2 N}_{i=0} w^i_p \left( f(\bar{x}^i_{k-1 \vert k-1}) - \hat{x}_{k \vert k-1} \right) \left( f(\bar{x}^i_{k-1 \vert k-1}) - \hat{x}_{k \vert k-1} \right)^T + Q_k
\end{equation}
\end{itemize}

{\bf Correction equations}

The correction equations take the form:
\begin{itemize}
\item Predicted measurements ($K$-dimensional ket)
\begin{equation} \label{eq:EQ036}
\hat{z}_k = \sum^{2 N}_{i=0} w^i_x h(\bar{y}^i_{k \vert k-1})
\end{equation}
\item Residuals ($K$-dimensional ket)
\begin{equation} \label{eq:EQ037}
\tilde{z}_k = z_k - \hat{z}_k
\end{equation}
\item Innovation matrix ($K \times K$ matrix)
\begin{equation} \label{eq:EQ038}
S_k = \sum^{2 N}_{i=0} w^i_p \left( h(\bar{y}^i_{k \vert k-1}) - \hat{z}_k \right) \left( h(\bar{y}^i_{k \vert k-1}) - \hat{z}_k \right)^T + R_k
\end{equation}
\item State-measurement cross-covariance matrix ($N \times K$ matrix)
\begin{equation} \label{eq:EQ039}
C_k = \sum^{2 N}_{i=0} w^i_p \left( f(\bar{x}^i_{k-1 \vert k-1}) - \hat{x}_{k \vert k-1} \right) \left( h(\bar{y}^i_{k \vert k-1}) - \hat{z}_k \right)^T
\end{equation}
\item Optimal gain ($N \times K$ matrix)
\begin{equation} \label{eq:EQ040}
K_k = C_k S_k^{-1}
\end{equation}
\item \emph{A posteriori} state estimate ($N$-dimensional ket)
\begin{equation} \label{eq:EQ041}
\hat{x}_{k \vert k} = \hat{x}_{k \vert k-1} + K_k \tilde{z}_k
\end{equation}
\item \emph{A posteriori} covariance matrix ($N \times N$ matrix)
\begin{equation} \label{eq:EQ042}
P_{k \vert k} = P_{k \vert k-1} - K_k S_k K_k^T
\end{equation}
\end{itemize}
The weights $w^i_x$ and $w^i_p$ (identical in both phases) have been detailed in Eqs.~(\ref{eq:EQ031}).

\subsection{\label{sec:UKFExample}Example of an application of the Unscented Kalman Filter}

The re-entry problem (i.e., the motion of an Earth-bound spacecraft, impacting the Earth's atmosphere), tailored to the case of the Space Shuttle, is described in Appendix \ref{App:AppB}. The position and 
the direction of motion of the incoming spacecraft are monitored by terrestrial radars, providing the measurements (for the filters). Mehra \cite{mehra} compared the performance of several non-linear filters 
on such data, including two EKFs (formulated in two different coordinate systems), whereas Chang, Whiting, and Athans \cite{caw} addressed the modelling accuracy and complexity, as well as the real-time 
processing requirements, also exploring techniques compensating for modelling inaccuracies by increasing the process noise. Austin and Leondes \cite{al} developed a robust filter, based on statistical 
linearisation. More recently, Crassidis and Markley \cite{cm}, as well as a number of other authors with contributions in the Proceedings of various AIAA and IEEE Conferences, dealt with the re-entry problem, 
which is generally regarded as stressful for filters and trackers \cite{ju1,ju2,mehra,al}.

\begin{figure}
\begin{center}
\includegraphics[width=15.5cm]{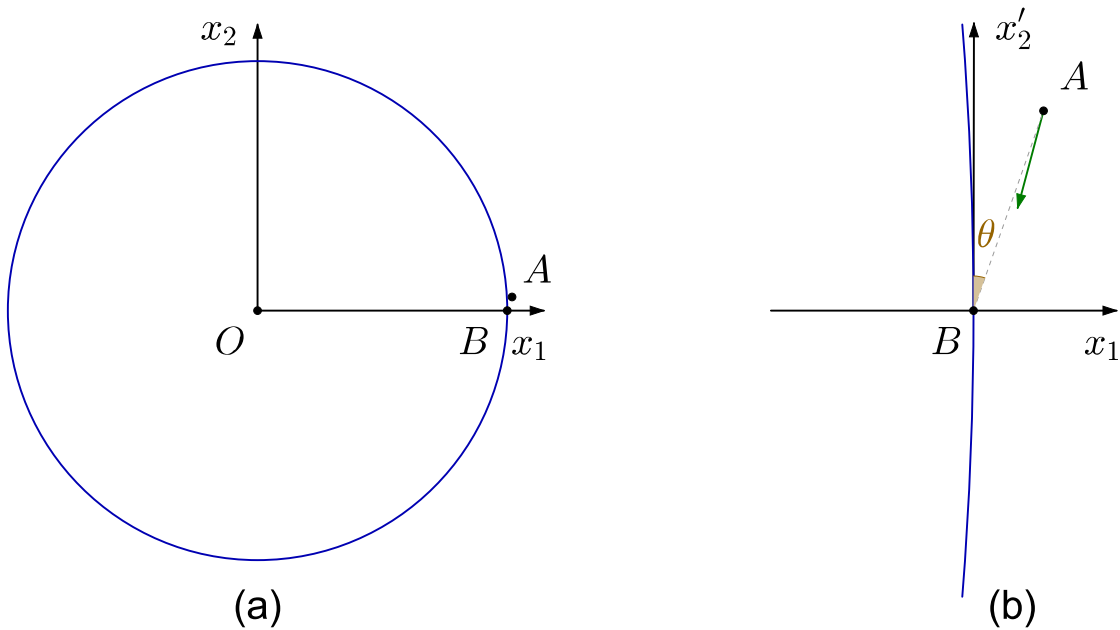}
\caption{\label{fig:ReentryGeometry}The coordinate system used in the re-entry problem. Point $A$ corresponds to the position of the vehicle at time $t=0$; as in Ref.~\cite{ju1}, an initial altitude of 
$131.632$ km is assumed. Figure (a), drawn to scale, demonstrates the smallness of this altitude in comparison to the radius of the Earth. Figure (b), also drawn to scale, is a close-up of Fig.~(a) around 
points $A$ and $B$; the axis $x^\prime_2$ is parallel to $x_2$. The green vector indicates the initial velocity of the vehicle ($-1.8093$ km/s,$-6.7967$ km/s) \cite{ju1}. The coordinate system ($x_1$,$x^\prime_2$), 
hence also ($x_1$,$x_2$), may be chosen at will; however, it makes sense to place the origin of the system in the vicinity of the region of interest, i.e., of the points $A$ and $B$.}
\vspace{0.35cm}
\end{center}
\end{figure}

One simplification of the problem is achieved by considering the $3$-dimensional motion of the vehicle as planar; in this approximation, the object remains on the plane defined by the velocity vector at 
the beginning of the re-entry and the centre of gravity, see Fig.~\ref{fig:ReentryGeometry}.

The dynamical model, followed in Ref.~\cite{ju1}, attempts the determination of the motion on the basis of three forces.
\begin{itemize}
\item The aerodynamic drag. This velocity- and altitude-dependent force is the dominant one.
\item The gravitational force.
\item Forces associated with random buffeting effects.
\end{itemize}
The state of the system is assumed to be a $5$-dimensional vector, comprising the coordinates $x_1$ and $x_2$ of the position vector, the corresponding velocities $x_3=\dot{x}_1$ and $x_4=\dot{x}_2$, as 
well as one term associated with the aerodynamic aspects of the motion ($x_5$). The rate of change of the two velocity components and of $x_5$, including the process noise $q$ of Eq.~(\ref{eq:EQ001}), is 
given in Ref.~\cite{ju1} as:
\begin{align} \label{eq:EQ043}
\dot{x}_3 &= A x_3 + B x_1 + q_3\nonumber\\
\dot{x}_4 &= A x_4 + B x_2 + q_4\nonumber\\
\dot{x}_5 &= q_5 \, \, \, ,
\end{align}
where
\begin{align} \label{eq:EQ044}
A &= -\gamma \exp \left( \frac{R - r}{r_c} \right) v\nonumber\\
B &= -\frac{G_N M}{r^3}\nonumber\\
\gamma &= \gamma_0 \exp(x_5)\nonumber\\
r &= \sqrt{x^2_1+x^2_2}\nonumber\\
v &= \sqrt{x^2_3+x^2_4} \, \, \, .
\end{align}
For the sake of brevity, all dependences on the appropriate variables, as well as on time, are not explicitly given in Eqs.~(\ref{eq:EQ043},\ref{eq:EQ044}). The constants are fixed as follows: $r_c=13.406$ 
km \cite{ju1}, $G_N=6.6738 \cdot 10^{-11}$ m$^3$ kg$^{-1}$ s$^{-2}$, $M=5.9726 \cdot 10^{24}$ kg, and $R=6\,378.137$ km \cite{pdg}.

Regarding this problem, the reader should be warned that Refs.~\cite{ju1,ju2} contain a number of mistakes; a subsequent short communication \cite{ju3} attempted corrections to some of these flaws, but did 
not cover all the issues raised below. The corrected flaws are as follows.
\begin{itemize}
\item A representative value of the ballistic coefficient $\gamma_0$ (this parameter being denoted as $\beta_0$ therein) was given in Refs.~\cite{ju1,ju2} as $-0.59783$ km$^{-1}$ (in both papers, most of 
the physical units were omitted; SI base units were generally assumed, save for the lengths which were expressed in km). However, given the overall sign of the function $A$ (denoted as $D$ therein), a 
negative $\gamma_0$ value would inevitably result in a drag force pointing toward the centre of gravity, i.e., in the direction of motion; therefore, when adopting the negative sign of Refs.~\cite{ju1,ju2} 
(see first of Eqs.~(\ref{eq:EQ044})), the correct $\gamma_0$ value is $+0.59783$ km$^{-1}$. Reference \cite{ju3} provided a correction by reverting the overall sign of the function $A$ (and retaining the 
$\gamma_0$ value given in Refs.~\cite{ju1,ju2}).
\item In the formula of the gravitational force, the quantity $r$ in Refs.~\cite{ju1,ju2} was replaced by $R$ \cite{ju3}.
\item The covariance matrix of the process noise $Q_k$, given in Refs.~\cite{ju1,ju2}, corresponds to the input in the Monte-Carlo simulation; the matrix driving the filter (to be given shortly) contains 
a non-zero diagonal element for the state-vector component $x_5$ \cite{ju3}.
\item Figures 5 \cite{ju1} and 9 \cite{ju2} are confusing. The caption of the figures refers to one solid line, yet two solid curves appear in each of these figures; the dots of the dotted curves cannot 
be discerned, and these curves also appear as solid (thicker, however, than the curves which were intended to be `solid lines'). Reference \cite{ju3} provided a set of improved figures.
\end{itemize}
For some inexplicable reason, the reviewers of these two works missed all these obvious problems. Two additional flaws in Refs.~\cite{ju1,ju2} remain uncorrected: one is an obvious `slip of the pen', 
whereas the second one is rather serious. Regarding the former, the statistical term `variance' is used in Section 4 of Ref.~\cite{ju1}, at the point where the covariance matrix of the measurement noise 
is introduced, rather than `standard deviation'; that passage was obviously copied-and-pasted to part B of Section VI of Ref.~\cite{ju2}, which contains the same mistake. The serious flaw will be discussed 
shortly.

Assumed in Ref.~\cite{ju1} is that the motion of the vehicle is monitored by a radar positioned at ($x_1$,$x_2$)=($x^r_1$,$x^r_2$); I will use $x^r_1=R$ and $x^r_2=0$, i.e., a radar positioned at point $B$ 
in Fig.~\ref{fig:ReentryGeometry}. The radar of Ref.~\cite{ju1} provides measurements of the distance $d$, as well as of the elevation angle $\theta$, at the sampling rate of $10$ Hz. Evidently,
\begin{align} \label{eq:EQ045}
d &= \sqrt{(x_1-x^r_1)^2+ (x_2-x^r_2)^2} + r_1\nonumber\\
\theta &= \arctan \left( \frac{x_2-x^r_2}{x_1-x^r_1} \right) + r_2 \, \, \, ,
\end{align}
where $r_1$ and $r_2$ represent the two components of the measurement noise; herein, the root-mean-square (rms) of the measurement noise was set to $1$ m for $d$ and $0.17$ mrad for $\theta$ \cite{caw}. 
In Fig.~\ref{fig:ReentryGeometry}, the distance $d$ of the first of Eqs.~(\ref{eq:EQ045}) is the magnitude of the position vector of the vehicle in the coordinate system ($x_1$,$x^\prime_2$) with origin 
at point $B$; $\theta$ is the angle of the position vector with the $x^\prime_2$ axis.

The second uncorrected flaw of Refs.~\cite{ju1,ju2} relates to the rms of the noise in the angle measurement. On page 102 of Ref.~\cite{caw}, Chang, Whiting, and Athans wrote in 1977: ``The angle measurement 
standard deviation is assumed to be $0.17$ mrad.'' Quoting that paper in Refs.~\cite{ju1,ju2}, Julier and Uhlmann explain: ``\dots $w_1(k)$ and $w_2(k)$ [equivalent to the quantities $r_1$ and $r_2$ of 
Eqs.~(\ref{eq:EQ045})] are zero mean uncorrelated noise processes with variances [\emph{sic}] of $1$ m and $17$ mrad, respectively.'' Up to the present time, it remains a mystery why the noise level in the 
measurement of the angle in Refs.~\cite{ju1,ju2} was increased by two orders of magnitude (that is, if the value quoted in Refs.~\cite{ju1,ju2} is not one additional mistype).

The actual state of the system at $t=0$ is given in Ref.~\cite{ju1} as
\begin{equation} \label{eq:EQ046}
x_0
= \left(
\begin{array}{c}
(6\,500.4000 \pm 0.0010) \text{ km}\\
(349.1400 \pm 0.0010) \text{ km}\\
(-1.8093 \pm 0.0010) \text{ km/s}\\
(-6.7967 \pm 0.0010) \text{ km/s}\\
0.6932
\end{array} \right)
\end{equation}
and the process noise as
\begin{equation} \label{eq:EQ047}
Q_k
= \left(
\begin{array}{ccccc}
0 & 0 & 0 & 0 & 0\\
0 & 0 & 0 & 0 & 0\\
0 & 0 & \sigma_3^2 & 0 & 0\\
0 & 0 & 0 & \sigma_4^2 & 0\\
0 & 0 & 0 & 0 & \sigma_5^2
\end{array} \right)
\end{equation}
where $\sigma^2_3=\sigma^2_4=2.4064 \cdot 10^{-5}$ km$^2$/s$^4$ \cite{ju1,ju2,ju3} and $\sigma_5^2=10^{-6}$ \cite{ju3}.

\begin{figure}
\begin{center}
\includegraphics[width=15.5cm]{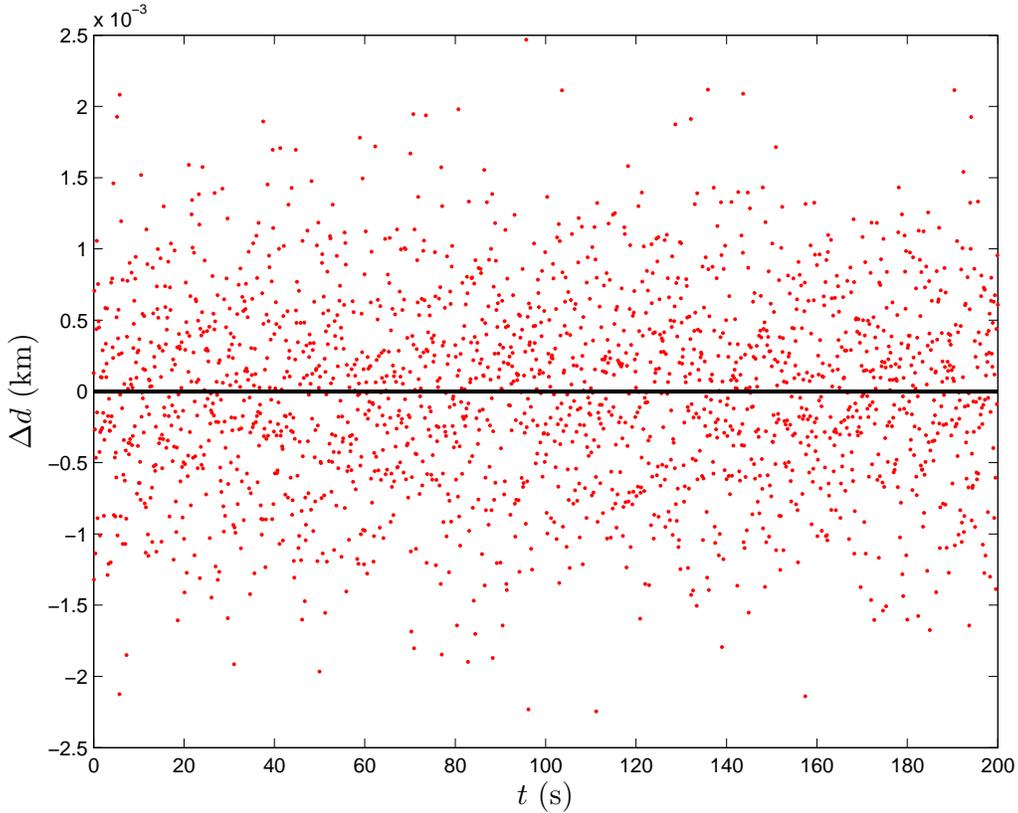}
\caption{\label{fig:ResidualsReentryDistance}The time series of the residuals between the simulated and filtered data of the distance $d$ for $0 \leq t \leq 200$ s.}
\vspace{0.35cm}
\end{center}
\end{figure}

\begin{figure}
\begin{center}
\includegraphics[width=15.5cm]{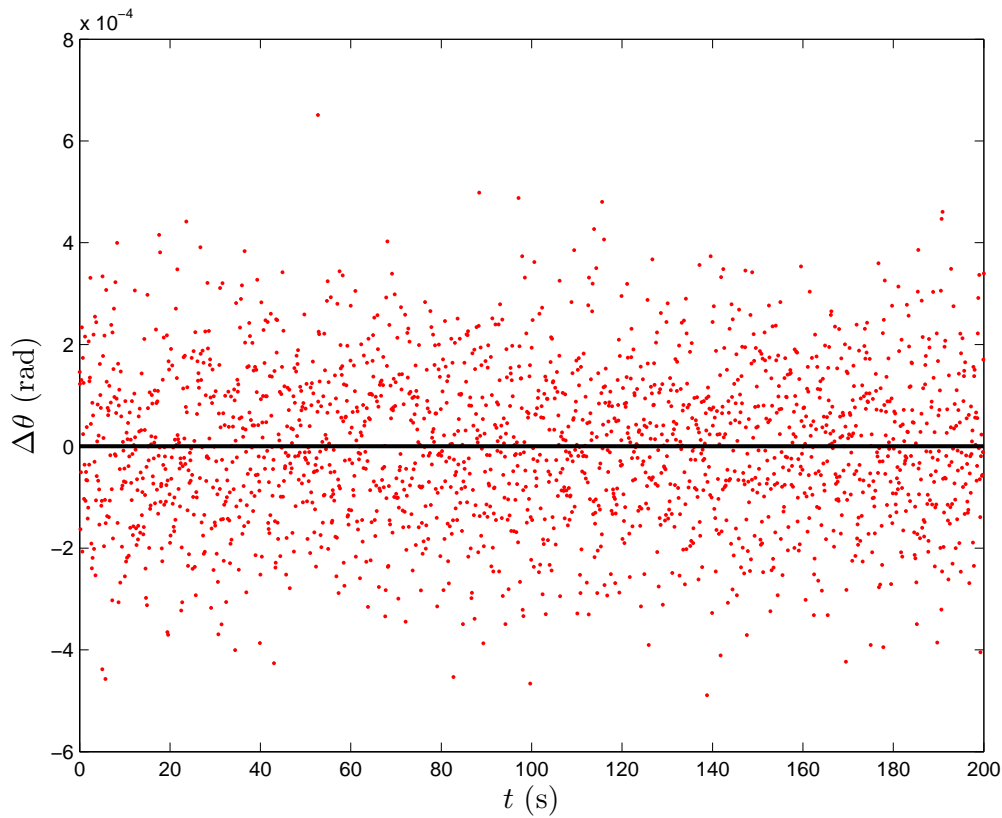}
\caption{\label{fig:ResidualsReentryAngle}The time series of the residuals between the simulated and filtered data of the elevation angle $\theta$ for $0 \leq t \leq 200$ s.}
\vspace{0.35cm}
\end{center}
\end{figure}

On the basis of this input, one may obtain series of simulated radar measurements ($d$,$\theta$) and filter the data using the UKF. One representative example of residuals (i.e., of the differences between 
the simulated and the filtered data) is displayed in Figs.~\ref{fig:ResidualsReentryDistance} and \ref{fig:ResidualsReentryAngle}. The reduced $\chi^2$ (i.e., the overall $\chi^2$ value divided by the number 
of degrees of freedom), obtained on the basis of these residuals for the measurement uncertainties of Ref.~\cite{caw}, comes out equal to about $0.66$. This value suggests that the filtering of the raw 
measurements is successful~\footnote{The reduced $\chi^2$ value is significantly smaller than $1$; this is the result of the obvious double-counting of the uncertainties in the simulation and in the processing 
of the simulated data.}. Furthermore, the residuals come out independent of the free variable in the problem (i.e., the time $t$); this is the expected result in all successful optimisation schemes. Plots 
\ref{fig:ResidualsReentryDistance} and \ref{fig:ResidualsReentryAngle} have been obtained using the $\alpha$, $\beta$, and $\kappa$ values of $10^{-3}$, $2$, and $0$, respectively \cite{wm}. It has been 
confirmed that the sensitivity of the present results to the variation of the parameters~\footnote{Kandepu, Foss, and Imsland \cite{kfi} favour larger $\alpha$ values (e.g., they use $\alpha=1$), as well as 
the choice $\lambda=3-N$ \cite{ju1}. In Ref.~\cite{sarkka}, S\"arkk\"a also used a `large' $\alpha$ value (namely, $0.5$); however, the comparison of his parameter values with those of other works might not 
be appropriate, as it is not clear whether the second of his Eqs.~(10) - i.e., the expression yielding his weight $W^{(c)}_0$, denoted as $w^0_p$ in this work, see the second of Eqs.~(\ref{eq:EQ031}) - is a 
mistype or represents the actual formula employed in S\"arkk\"a's work.} $\alpha$ (fixed at $10^{-4}$, $10^{-3}$, $0.1$, $0.5$, and $1$) and $\kappa$ (fixed at $-2$ and $0$) is very low. For $\alpha \geq 10^{-3}$, 
the reduced $\chi^2$ varied between $0.66008$ and $0.66016$; somewhat smaller values were obtained for $\alpha = 10^{-4}$, but the temporal dependence of the diagonal elements of the covariance matrix $P$ 
was rather noisy.

The effort notwithstanding, I have not been able to reproduce the temporal dependence of the variance of the state-vector components $x_1$, $x_3$, and $x_5$, given in Refs.~\cite{ju1,ju2,ju3}; unfortunately, 
it has not been possible to resolve this issue with the first author of these works. The inability to reproduce these plots may be due to my misinterpretation of the process and measurement covariance 
matrices, given in these three papers. S\"arkk\"a, who also studied the re-entry problem in Ref.~\cite{sarkka}, did not report discrepancies with the results of Refs.~\cite{ju1,ju2,ju3}.

The figures in this paper have been produced either with CaRMetal, a dynamic geometry free software (GNU-GPL license), first developed by Ren\'e Grothmann and recently under Eric Hakenholz \cite{CMl}, or 
with MATLAB.

\newpage
\appendix
\section{\label{App:AppA}Motion of a massive object in the gravitational field of the Earth taking account of the air resistance}

The equation for viscous resistance is applicable in the case of slow motion of an object through a fluid; the turbulence may be neglected and the force, associated with the resistance of the fluid, 
is proportional to the velocity of the object:
\begin{equation} \label{eq:EQA01}
\vec{F} = - b \vec{v} \, \, \, ,
\end{equation}
where $b>0$ is a constant depending on the fluid and on geometrical properties of the object in motion.

Taking account of the air resistance of Eq.~(\ref{eq:EQA01}), Newton's second law of motion in case of a massive object, moving in the gravitational field of the Earth, is of the form:
\begin{equation} \label{eq:EQA02}
m \frac{d^2x}{dt^2} = - m g - b \frac{dx}{dt} \, \, \, ,
\end{equation}
where $m$ denotes the mass of the object. This equation may be written as:
\begin{equation} \label{eq:EQA03}
\frac{d^2x}{dt^2} = - g - \mu \frac{dx}{dt} \Rightarrow - \frac{dv}{g + \mu v} = dt \, \, \, ,
\end{equation}
where $\mu=b/m$. The solution of this equation, compatible with the initial condition $v(0)=v_0$, is:
\begin{equation} \label{eq:EQA04}
v(t) = v_0 e^{-\mu t} - \frac{g}{\mu} (1 - e^{-\mu t}) \, \, \, .
\end{equation}
For $\mu t \ll 1$, $1 - e^{-\mu t} \approx \mu t$ and Eq.~(\ref{eq:EQ021}) is retrieved.

The altitude $x(t)$ is easily obtained via the integration of Eq.~(\ref{eq:EQA04}). The solution, compatible with the initial condition $x(0)=x_0$, is:
\begin{equation} \label{eq:EQA05}
x(t) = x_0 + \frac{v_0}{\mu} (1 - e^{-\mu t}) + \frac{g}{\mu^2} (1 - e^{-\mu t}) - \frac{gt}{\mu} \, \, \, .
\end{equation}
For $\mu t \ll 1$, it is easy to show that Eq.~(\ref{eq:EQ020}) is retrieved. As the square of the quantity $\mu$ appears in the denominator of the third term on the right-hand side of 
Eq.~(\ref{eq:EQA05}), it is necessary to retain (in that part) the terms up to $\mu^2 t^2$ in the Taylor expansion of $e^{-\mu t}$; for $\mu t \ll 1$, $1 - e^{-\mu t} \approx \mu t - \mu^2 t^2/2$.

\section{\label{App:AppB}The re-entry problem}

The re-entry procedure of the Space Shuttle (henceforth, simply `Orbiter') is initiated at an altitude of $150$ km (above the surface of the Earth), slightly over one hour before touchdown. At this altitude, 
the velocity of the circular orbit is about $7.81$ km/s. The Orbiter revolves around the Earth upside-down, its top side facing toward the Globe, with its nose in the direction of the motion. The Orbiter is 
first flipped about its pitch axis via the application of thrusters. The OMS (Orbital Maneuvering System) engines face now forward and their engagement provides the necessary decrease in velocity (by slightly 
over $300$ km/h), bringing the orbit's perigee into the upper part of the atmosphere and committing the Orbiter to the descent. The Orbiter is next flipped - via lateral RCS (Reaction Control System) 
thrusters - about its yaw axis and attains descent configuration, its nose being in the general direction of the motion and diving toward the surface of the Earth.

At an altitude of about $120$ km, the hypersonic Orbiter enters the upper part of the atmosphere. It is critical that the angle at which the Orbiter impinges onto the atmosphere (also referred to as the 
`angle of attack') be kept within a narrow margin of $40^\circ$, thus restricting the motion of the Orbiter within a $3$-dimensional re-entry corridor; significantly smaller angles would make the Orbiter 
`bounce off' the atmosphere, whereas larger ones would result in its incineration and the loss of the crew.

During the re-entry, the kinetic energy of the Orbiter is converted into heat and emitted into the atmosphere. In the upper part of the atmosphere, where the air is tenuous, the descent generates a pressure 
wave, compressing the air ahead of the plunging Orbiter and causing intense atmospheric ionisation. The largest amount of heat, which the Orbiter suffers during re-entry, is dissipated via this detached 
shock wave. Viscous drag becomes progressively more important as the air density rises with decreasing altitude.

From the moment when the re-entry procedure is initiated to an altitude of about $85$ km, the motion of the Orbiter is ballistic. At about this altitude, the Orbiter becomes manoeuvrable, turning into a 
giant (and inelegant) glider. To expedite the loss of kinetic energy, the Orbiter undergoes four steep S-shaped banking turns. In terms of resilience to heat and efficiency in heat dissipation, this is the 
most demanding phase of the re-entry. Temperatures rise up to $1650$ $^\circ$C; the only protection of the Orbiter and of its crew consists of the surface-insulation tiles on the Orbiter's underside, a 
$4.4$-tonne wall between life and death.

The drag on the Orbiter increases considerably as it comes into the thicker parts of the atmosphere, the Orbiter decelerates at increasing rate, and the autopilot disengages. Manual control of the Orbiter 
is now assumed and its heading is adjusted for landing. Shortly before touchdown, the velocity of the Orbiter drops below the speed of sound. As the Orbiter is aligned with the runway, it is set into 
a steep descent, pitching at almost $20^\circ$ below the local horizontal, seven times steeper than commercial airplanes.

\end{document}